\documentclass{moriond}

\def\be{\begin{equation}}
\def\ee{\end{equation}}
\def\bea{\begin{eqnarray}}
\def\eea{\end{eqnarray}}

\newcommand{\ttbar}{\ensuremath{\mathrm{t\bar{t}}}}

\newcommand{\mttbar}{\ensuremath{m_\mathrm{{t\bar{t}}}}}
\newcommand{\mt}{\ensuremath{\mathrm{m_{t}}}}

\newcommand{\Photo}{}

\begin{document}
\vspace*{4cm}
\title{Top quark properties measurements and searches at ATLAS and CMS}

\author{Sebastian Wuchterl\\on behalf of the ATLAS and CMS Collaborations\footnote{Copyright [2024] CERN for the benefit of the ATLAS and CMS Collaborations. Reproduction of this article or parts of it is allowed as specified in the CC-BY-4.0 license}}

\address{CERN, Espl. des Particules 1,\\
1211 Meyrin, Switzerland}

\maketitle\abstracts{
Theoretical predictions for processes involving top quarks, such as top quark-antiquark pair production, depend on fundamental standard model (SM) parameters like the top quark mass or potentially receive contributions from beyond-the-SM (BSM) physics. By comparing predictions with measurements performed at the ATLAS and CMS experiments, SM parameters can be determined precisely, or measurements can be used as a powerful probe for BSM physics. In these proceedings, recent results measuring SM and top quark properties are reported, as well as new results of searches for BSM physics for charged lepton flavor violation, flavor changing neutral currents, and baryon number violation are presented.}

\section{Introduction}
The standard model (SM) of particle physics yields excellent agreement with the vast majority of experimental measurements yet fails to explain phenomena such as the matter-antimatter asymmetry in the universe, dark matter, or neutrino masses. Thus, the precision study of its parameters, measurements of processes accessible at the energy scale of the CERN Large Hadron Collider (LHC), and searches for beyond-the-SM (BSM) physics are of crucial importance.
The top quark with a mass of about 172 GeV is the most massive particle of the SM, which implies that the top quark plays a key role within the SM, and processes involving top quarks receive potentially large contributions due to BSM physics.
In the SM, top quarks are predominantly produced in top-quark-antiquark pair (\ttbar) production, but with large data sets available after the second data-taking period (Run 2) at the LHC, also processes such as single top quark production become accessible for precise measurements.
It is therefore of the highest importance to enhance our knowledge of the SM via precision property measurements but also search for interactions forbidden in the SM at the tree level, such as charged lepton flavor violation (cLFV), baryon number violation (BNV), or flavor changing neutral currents (FCNC).
These proceedings report on the most recent results of the ATLAS~\cite{bib:ATLAS} and CMS~\cite{bib:CMS} Collaborations.

\section{Standard model properties measurements}
The value of the top quark mass is measured in different ways, and its definition depends on the experimental procedure employed or the theoretical definition used. The majority of measurements are performed via comparisons of detector-level distributions sensitive to the value of the mass to multi-purpose Monte-Carlo (MC) simulated templates. These types of measurements are commonly denoted as direct measurements whereas indirect measurements extract the top quark mass through comparisons of absolute or differential cross-section measurements at particle and parton levels to fixed-order theoretical predictions. Top quark mass measurements of the first type can be related to the top quark mass as defined in the pole-mass scheme (with an additional interpretation uncertainty), and measurements of the second kind directly extract the mass value for a given renormalization scheme.

Both the ATLAS and CMS experiments have performed a multitude of direct measurements during the LHC Run 1 from 2011--2012 in several production modes and final states at the center of mass energies of 7 and 8 TeV. In a recent combination~\cite{bib:MassCombination}, 15 individual measurements are combined, of which six are performed by ATLAS and nine by CMS experiment, respectively. The input measurements comprise analyses in the \ttbar\ and single top modes, as well as alternative methods using, e.g., soft leptons from decay products of the top quark.

\begin{figure}
\begin{minipage}{0.45\linewidth}
\centerline{\includegraphics[width=0.99\linewidth]{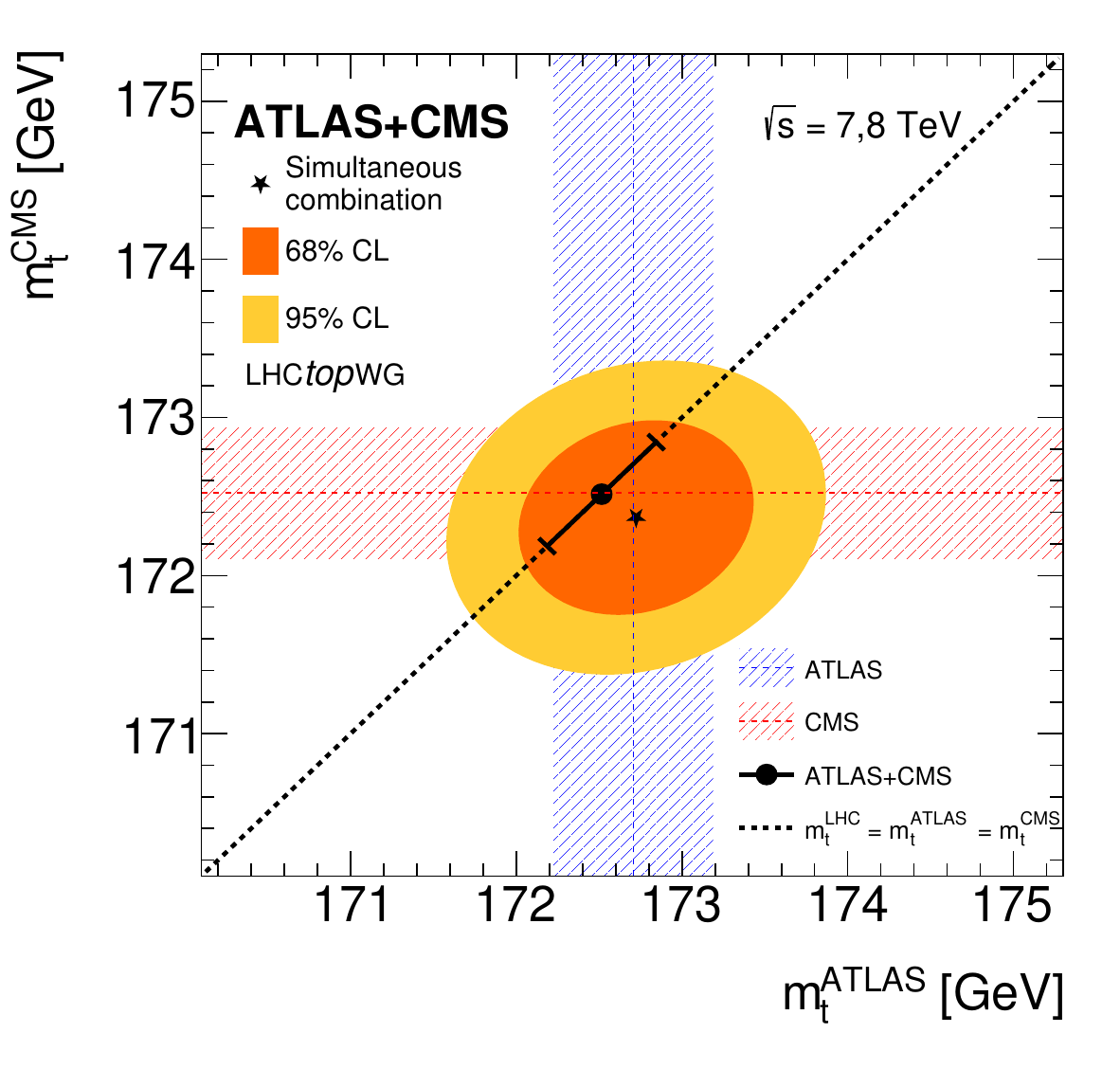}}
\end{minipage}
\hfill
\begin{minipage}{0.54\linewidth}
\centerline{\includegraphics[width=0.99\linewidth]{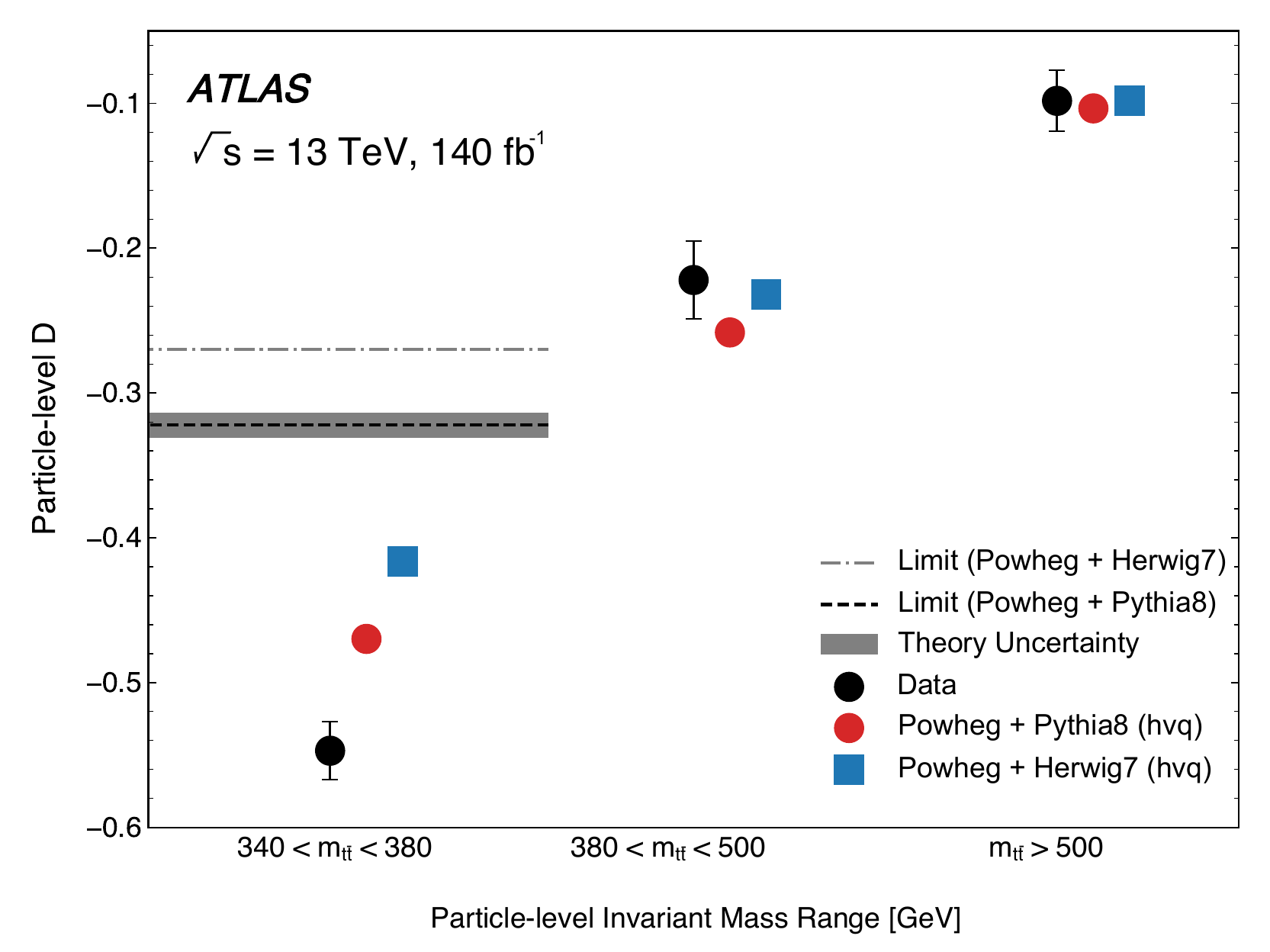}}
\end{minipage}
\caption[]{LHC top quark mass combination, with the individual ATLAS and CMS experiment combined values (left)~\cite{bib:MassCombination}. The entanglement measurement results in different \mttbar\ bins as performed by the ATLAS Collaboration~\cite{bib:ATLASEntanglement}.}
\label{fig:properties1}
\end{figure}

The combination is performed using the best linear unbiased estimator (BLUE) method~\cite{bib:BLUE}, where a weight ranging from zero to one is assigned to every input measurement, and their sum equals unity. In addition to the overall LHC combination, individual ATLAS and CMS experiment combined results are obtained.
A detailed study of correlations between the 15 measurements is performed, and the impact of the chosen correlation parameter for each uncertainty is evaluated on the final result and its uncertainty. Most of the correlation parameters are found to have minor or negligible impact, and the result is stable for the chosen assumptions.
The left plot in Fig.~\ref{fig:properties1} shows the three combined results. The LHC final result is $ \mt = 172.5 \pm 0.14\,(\mathrm{stat.}) \pm 0.30\,(\mathrm{syst.})\,\mathrm{GeV}$,
which improves 31\% compared to the most precise input measurement of the CMS experiment. The ATLAS and CMS experiment combined results, also shown in Fig.\ref{fig:properties1}, are in very good agreement with the total combination.
The final result shows a precision comparable to the most precise measurements of Run 2~\cite{bib:massljets}, which are discussed in a recent summary of \mt\ measurements of the CMS~\cite{bib:MassReview} Collaboration.

As the lifetime of top quarks are on the order of $10^{-25}\mathrm{s}$ and the hadronization time ($10^{-24}\mathrm{s}$) and spin-decorrelation time ($10^{-21}\mathrm{s}$) are significantly smaller, the decay products carry the spin information of the top quarks and provide a unique tool to study spin-correlations and correlated effects at the highest energies.
The \ttbar\ system can be described as a two-qubit system, and hence via the study of \ttbar\ production in the gluon-gluon fusion production mode where the quark pairs are produced in a spin singlet, the quantum entanglement of top quark pairs can be probed. As the differential cross section as a function of the angle $\phi$ between the two decay leptons in the final state events in the \ttbar\ rest system depends on the spin-correlation matrix, an entanglement marker of the form $D<-1/3$ can be defined, and $D$ can be extracted from $D = -\langle3 \cdot \cos{\phi}\rangle$.

In a recent ATLAS Collaboration measurement~\cite{bib:ATLASEntanglement}, for the first time at the LHC, quantum entanglement is probed with high significance. The measurement is performed using high-purity electron-muon events, and the signal region is defined for invariant \ttbar\ masses of $340\,\mathrm{GeV} < \mttbar < 380\,\mathrm{GeV}$ where the top quarks are found to be maximally entangled. Using a calibration curve, the entanglement marker $D$ is directly extracted at the particle level and is measured to be $D = -0.547 \pm 0.002\,(\mathrm{stat.}) \pm 0.021\,(\mathrm{syst.})$, well below the entanglement criterion of $1/3$.

\section{Searches for new physics}

\begin{figure}
\begin{minipage}{0.4\linewidth}
\centerline{\includegraphics[width=0.95\linewidth]{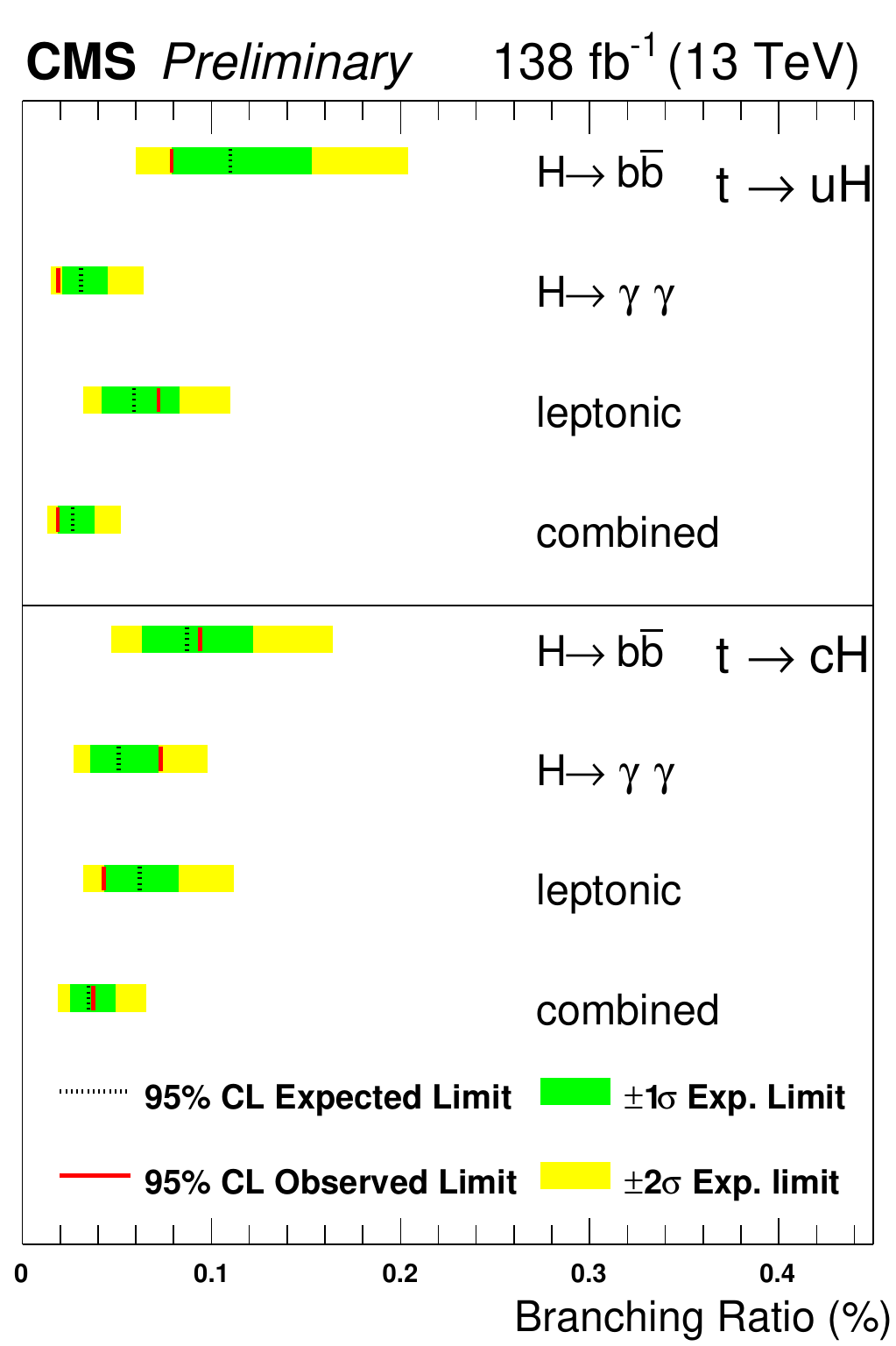}}
\end{minipage}
\hfill
\begin{minipage}{0.49\linewidth}
\centerline{\includegraphics[width=0.8\linewidth]{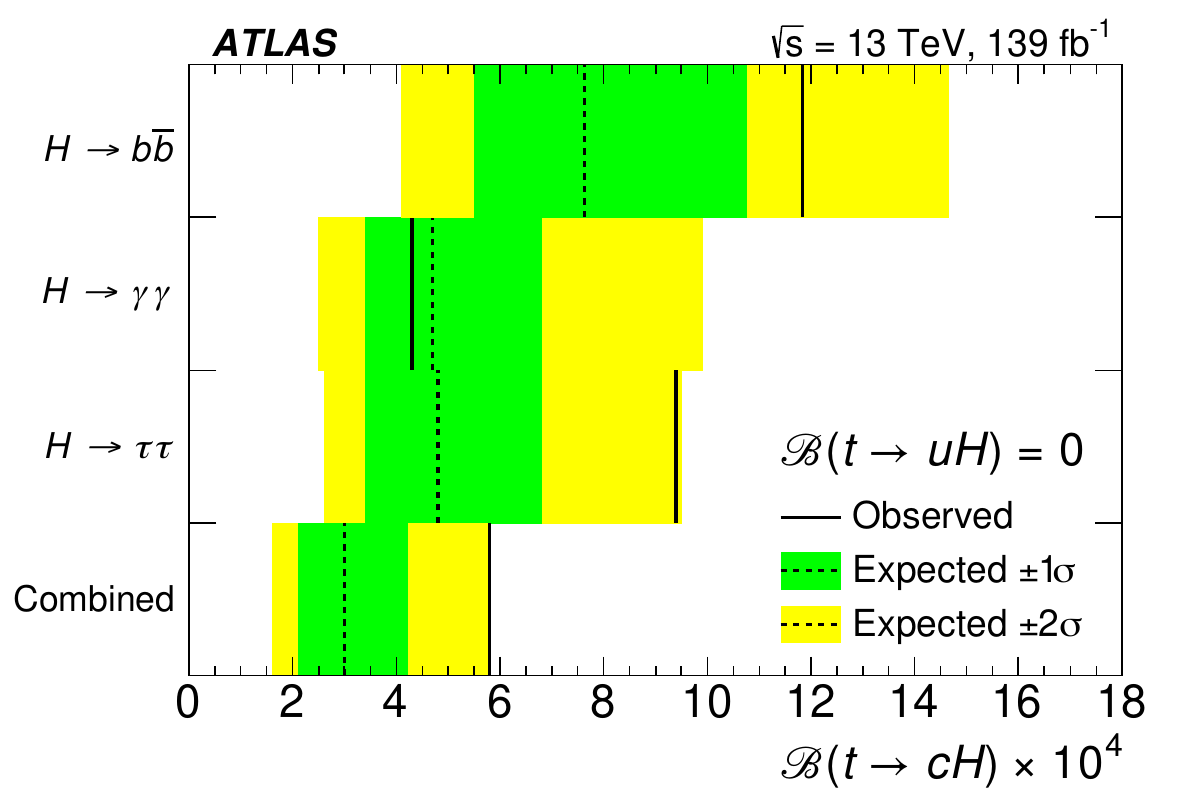}}
\centerline{\includegraphics[width=0.8\linewidth]{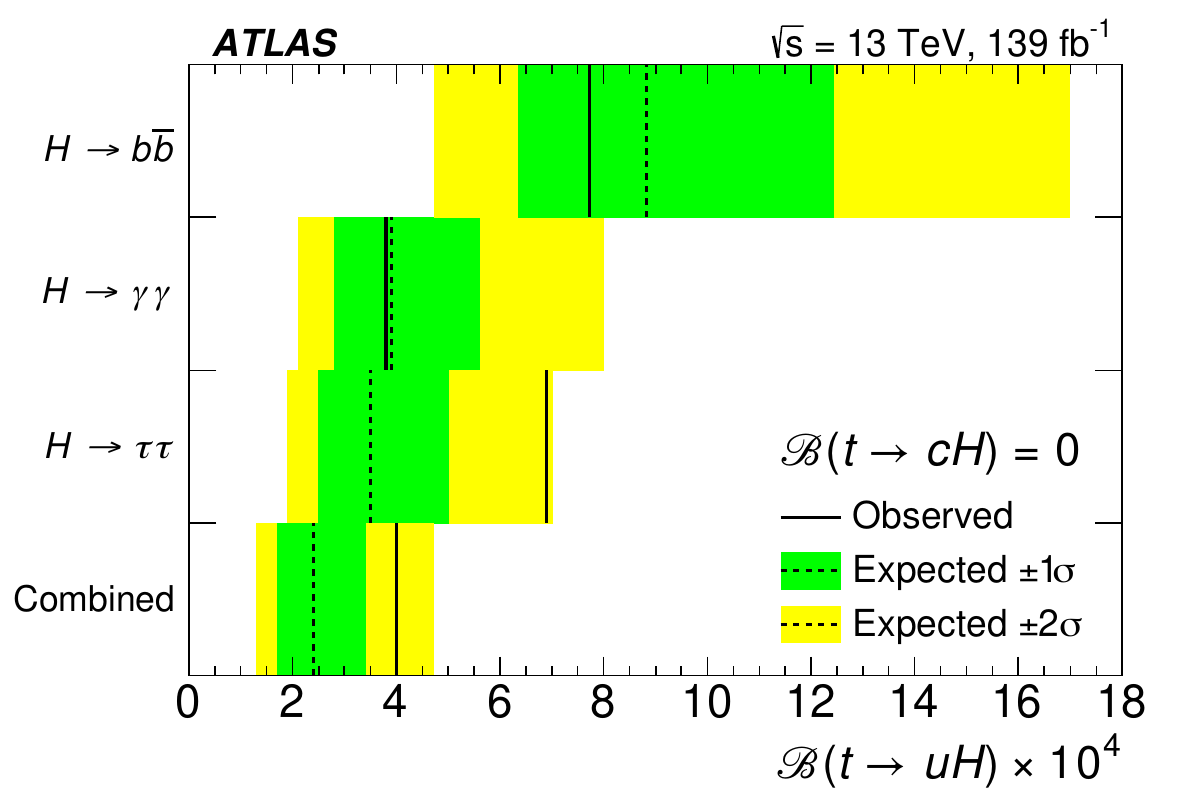}}
\end{minipage}
\caption[]{Upper limits on FCNC branching ratios as obtained by the CMS~\cite{bib:CMSFCNC} (left) and ATLAS experiment measurements~\cite{bib:ATLASFCNC} (right).}
\label{fig:searches1}
\end{figure}

Processes involving FCNCs are forbidden at tree level in the SM, but their enhancement would be direct evidence of BSM physics. The CMS and ATLAS Collaborations have reported new results on searches for FCNC in top-Higgs-associated production and the decay of one of the top quarks ($\mathrm{t}\rightarrow \mathrm{qH}$) in \ttbar\ production~\cite{bib:CMSFCNC,bib:ATLASFCNC}. The coupling to the Higgs boson is probed as it is expected to be significantly enhanced in multiple BSM scenarios such as two Higgs doublet models.
The measurement of the CMS experiment searches for FCNC in same-sign dilepton events, whereas the ATLAS experiment measurement analyses events with two photons in the final state. Both measurements made use of multivariate methods in the form of boosted decision trees (BDTs) to enhance the sensitivity. The analysis of the CMS Collaboration yields an upper limit on the branching ratios ($\mathcal{B}$) $\mathcal{B}(\mathrm{t}\rightarrow \mathrm{uH})$ and $\mathcal{B}(\mathrm{t}\rightarrow \mathrm{cH})$ of $0.072\%$ and $0.043\%$ at 95\% confidence level, respectively. The ATLAS experiment reports an upper limit of $0.038\%$ and $0.043\%$.
Both analyses also combine the results with previous measurements in different final states. The dilepton result of the CMS experiment is combined with measurements in the diphoton and bottom-quark-antiquark final states~\cite{bib:CMSFCNC_bb,bib:CMSFCNC_yy}. The combined result yields an upper limit of $0.037\%$ ($0.035\%$) for $\mathcal{B}(\mathrm{t}\rightarrow \mathrm{uH})$ ($\mathcal{B}(\mathrm{t}\rightarrow \mathrm{cH})$).
In the case of ATLAS experiment, the combined values are $0.040\%$ ($0.058\%$) for $\mathcal{B}(\mathrm{t}\rightarrow \mathrm{uH})$ ($\mathcal{B}(\mathrm{t}\rightarrow \mathrm{cH})$), using results from searches in the bottom-quark-antiquark~\cite{bib:ATLASFCNC_bb} and di-tau~\cite{bib:ATLASFCNC_tautau} final states.

\begin{figure}
\begin{minipage}{0.3\linewidth}
\centerline{\includegraphics[width=0.95\linewidth]{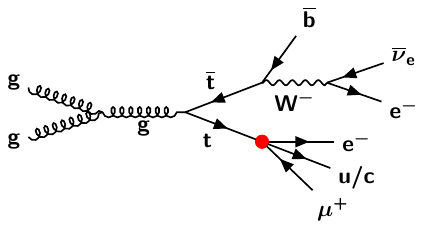}}
\end{minipage}
\begin{minipage}{0.3\linewidth}
\centerline{\includegraphics[width=0.95\linewidth]{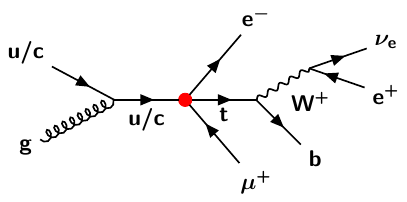}}
\end{minipage}
\begin{minipage}{0.3\linewidth}
\centerline{\includegraphics[width=0.95\linewidth]{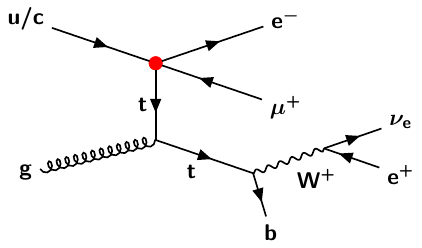}}
\end{minipage}
\hfill
\newline
\hfill
\begin{minipage}{0.99\linewidth}\centering
\centerline{\includegraphics[width=0.6\linewidth]{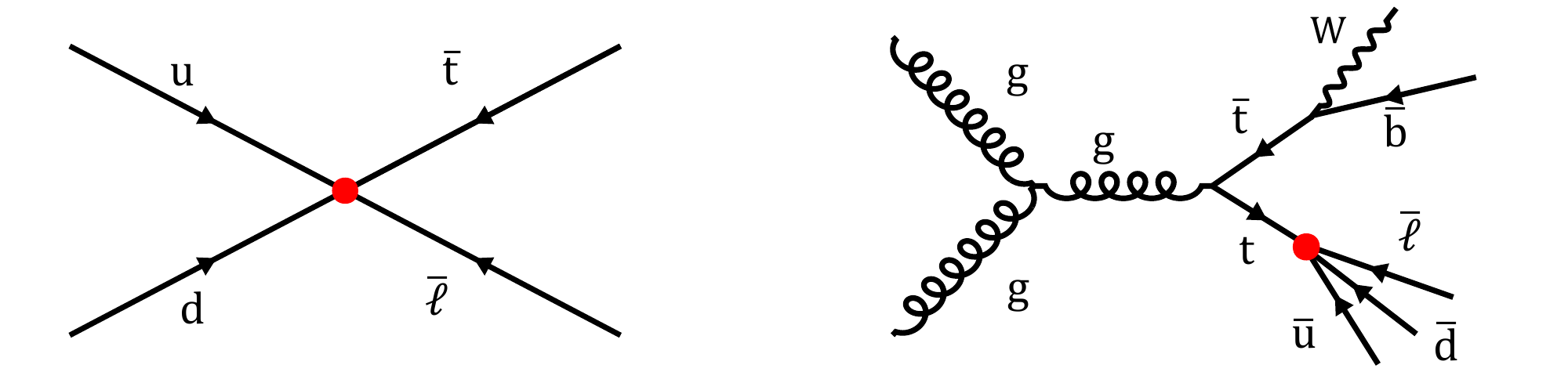}}
\end{minipage}
\caption[]{Sketched diagrams for FCNC ~\cite{bib:CMSFCNC} (top) and cLFV interactions~\cite{bib:CMSlfv} (bottom) involving top quarks.}
\label{fig:searches2}
\end{figure}

\begin{figure}
\centering
\begin{minipage}{0.8\linewidth}
\centerline{\includegraphics[width=0.99\linewidth]{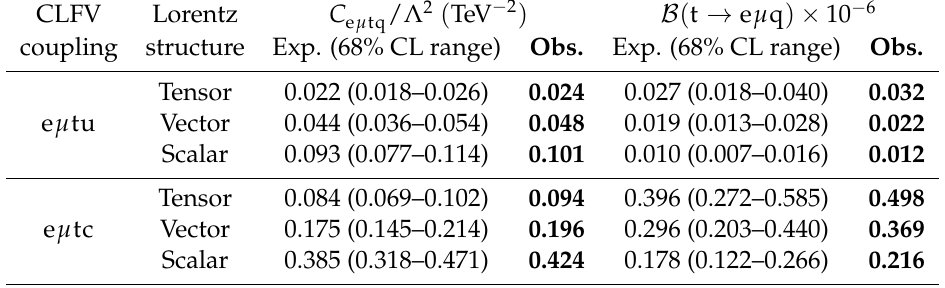}}
\end{minipage}
\hfill
\caption[]{Upper limits on the coupling coefficients and branching fractions as resulting from the cLFV analysis as performed by the CMS experiment~\cite{bib:CMSlfv}.}
\label{fig:searches3}
\end{figure}

Lepton flavor is a conserved quantity in the SM, but the existence of neutrino oscillations hints that there might also be a highly suppressed contribution of cLFV in the SM. Using processes involving top quarks, similar to the FCNC searches, this effect can be probed. Three possible diagrams for cLFV are shown in Fig.~\ref{fig:searches2} (top).
A recent analysis by the CMS experiment~\cite{bib:CMSlfv} searched for cLFV in the production and decay of top quarks, analyzing events with three leptons in the final state, where combinations of electrons and muons are considered. BDTs are used to enhance the sensitivity to the signal, which is parametrized in an effective field theory (EFT) approach using three independent operators (tensor, vector, and scalar interactions) contributing to four-fermion operators. Limits on these operators can be translated into limits on branching ratios for charged flavor-violating decays of the top quark due to their dependence on $\mathcal{B}$ and the production cross section.
The resulting expected and observed upper limits on the coupling coefficients in EFT and $\mathcal{B}$ for the different decay possibilities are given in the table in Fig.~\ref{fig:searches3}. Compared to previous analyses, the limits improve by about a factor of ten for all decay channels.

\begin{figure}
\centering
\begin{minipage}{0.49\linewidth}
\centerline{\includegraphics[width=0.99\linewidth]{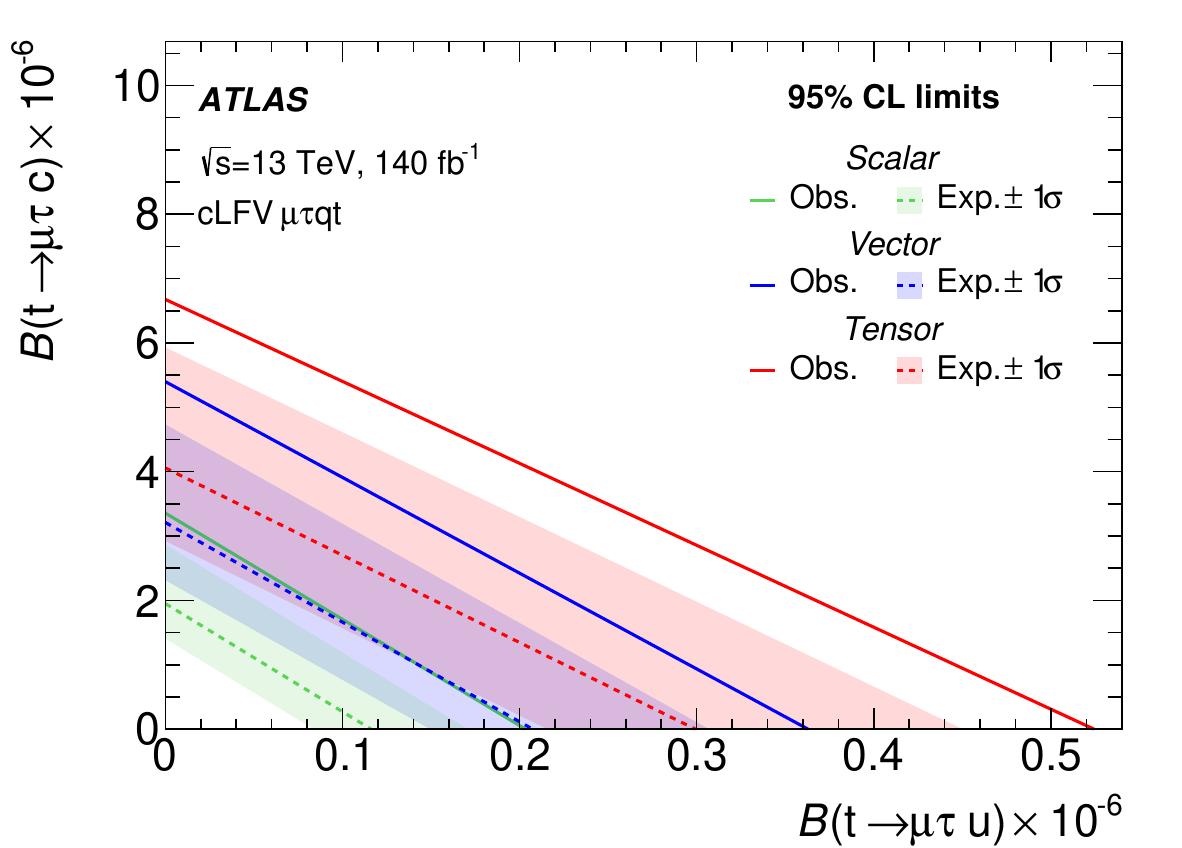}}
\end{minipage}
\begin{minipage}{0.49\linewidth}
\centerline{\includegraphics[width=0.99\linewidth]{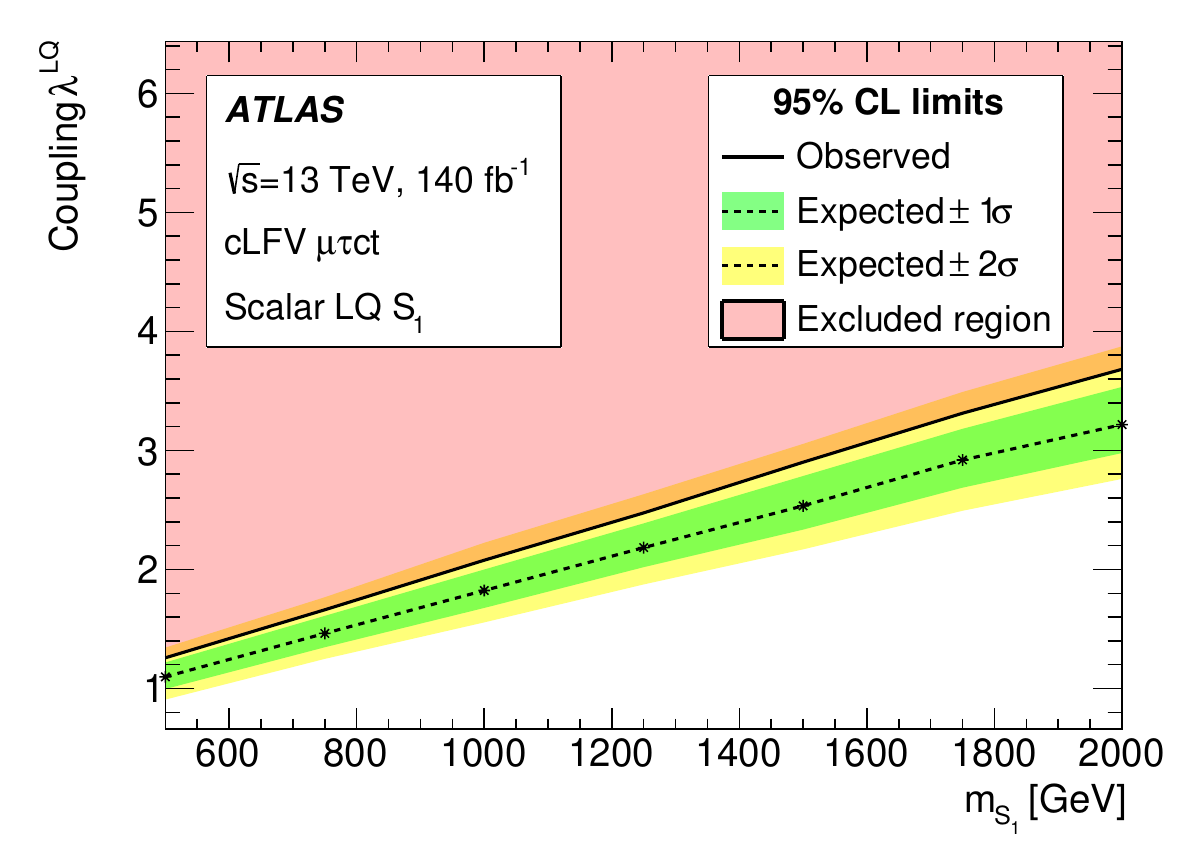}}
\end{minipage}
\hfill
\caption[]{Upper limits on the branching fractions of cLFV interactions involving top quarks (left) from the search as performed by the ATLAS Collaboration~\cite{bib:ATLASlfv} and the leptoquark interpretation (right).}
\label{fig:searches4}
\end{figure}

A similar analysis is carried out by the ATLAS experiment, targeting the same cLFV interactions in production and decay topologies~\cite{bib:ATLASlfv}. This search focuses on final state events containing one muon and one tau lepton, which decays hadronically, and is the first search for cLFV with tau leptons in the final state. The results are shown in Fig.~\ref{fig:searches4} in the left plot. An improvement on the same scale as for the ATLAS analysis on the profiled interactions to a previous result based on a reinterpretation of another measurement probing the same operators is observed.
In addition, the analysis interprets the results in a scalar leptoquark model, as leptoquarks are naturally able to serve as a source of cLFV. The considered model adds one scalar leptoquark $\mathrm{S_1}$ coupling to all up-type quarks and charged leptons. Upper limits in the two-dimensional plane of coupling versus $\mathrm{S_1}$ mass are shown in Fig.~\ref{fig:searches4} right.

\begin{figure}
\centering
\begin{minipage}{0.49\linewidth}
\centerline{\includegraphics[width=0.95\linewidth]{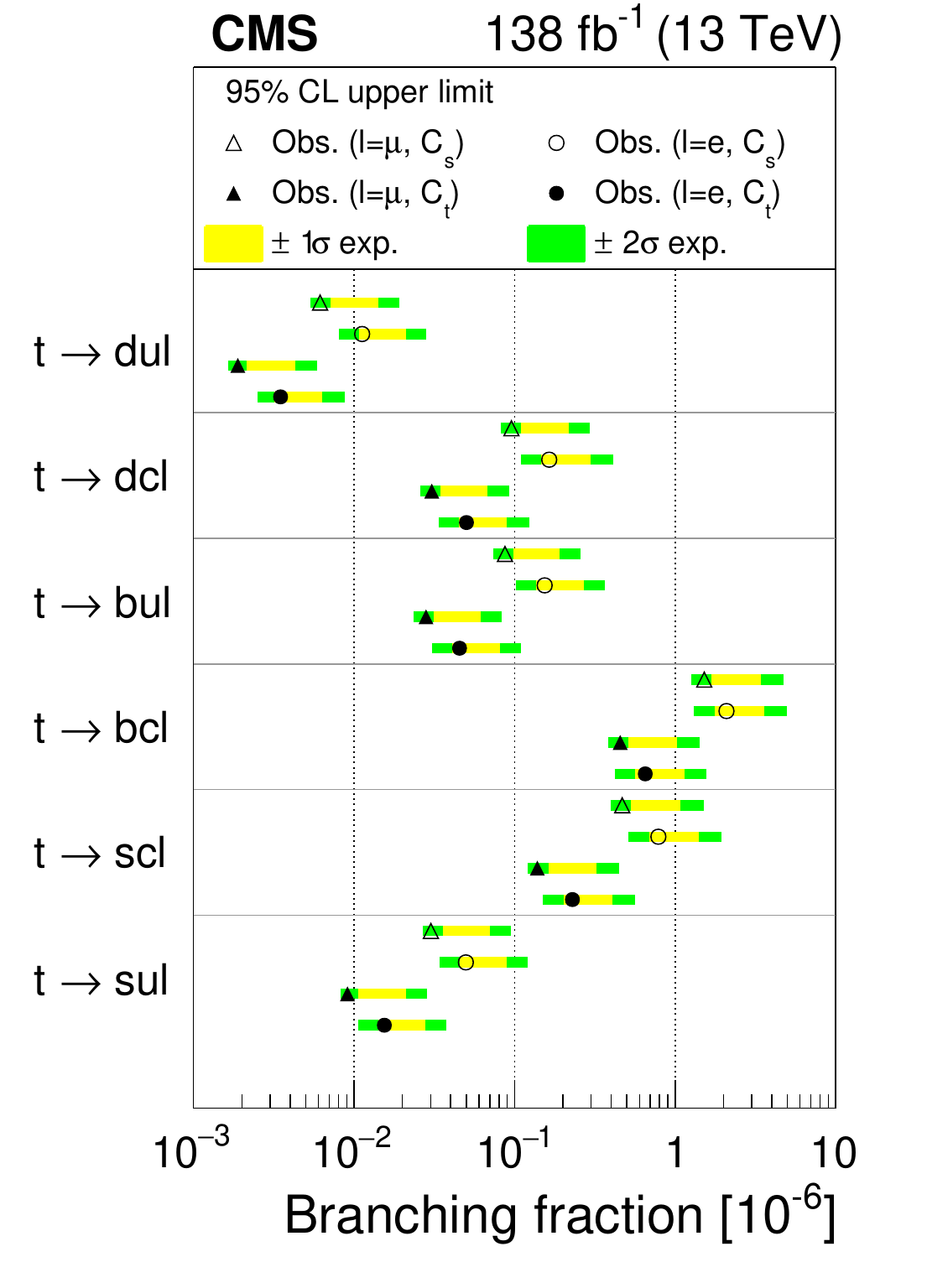}}
\end{minipage}
\hfill
\caption[]{Upper limits on the branching fractions involving BNV interactions with top quarks~\cite{bib:bnv}.}
\label{fig:searches5}
\end{figure}

Finally, the following analysis by the CMS Collaboration probes BNV in processes involving top quarks~\cite{bib:bnv}. To explain the matter-antimatter asymmetry in the universe, a significant amount of BNV is needed, although in the SM the baryon number is a conserved quantum number.
The search is performed in a model-independent way, also employing the EFT approach. Possible diagrams leading to BNV are displayed at the bottom in Fig.~\ref{fig:searches2}.
Both contributions from single top quark production as well as in the top quark decay are considered.
One combined BDT is used to discriminate between the background and BNV signal. The extracted expected and observed upper limits on baryon-number violating branchings are shown in Fig.~\ref{fig:searches5} for all contributing top quark decays.
Compared to previous results, these limits improve by multiple orders of magnitude, making these the most stringent limits to date.

\section{Conclusions}
Measurements of standard model (SM) processes involving top quarks are a powerful tool to extract SM parameters such as the top quark mass but also serve as an important tool to probe fundamental effects such as quantum entanglement or test the SM at many different fronts. Searches for beyond-the-SM phenomena such as quantum number violating interactions or suppressed processes. The ATLAS and CMS Collaborations have performed several measurements and searches with promising results, and the combination of individual results shows significant improvement with respect to the inputs.
Finally, the upcoming and ongoing Run 3 LHC data taking and even improved analysis techniques give rise to possibilities to observe rare processes and increase the precision of important SM measurements.

\section*{References}
\bibliography{references}{}
\end{document}